\documentclass[prd,twocolumn,floatfix,nofootinbib,amsmath,amssymb,floatfix]{revtex4} 
\usepackage{graphicx,color,dcolumn,booktabs,bm,mathrsfs}
\usepackage{longtable,lscape}
\usepackage{txfonts}
\usepackage{overpic}
\usepackage{amssymb}
\usepackage{indentfirst}
\usepackage{feynmf}   
\usepackage{slashed}  
\usepackage{cases}
\usepackage{color}			
\usepackage{multirow}
\usepackage{epstopdf}
\usepackage{CJK}

\def\tcsbar{{T_{c\bar{s}0}(2900)}}
\def\proc{{B^- \to K^- D^0 K^0}}
\def\dzkz{{D^0 K^0}}
\usepackage[colorlinks,
            citecolor=blue,
            anchorcolor=red,
            menucolor=red,
            linkcolor=red,
            filecolor=red,
            runcolor=red,
            urlcolor=blue,
            frenchlinks=red]{hyperref}
\usepackage{subfigure}

\begin{document}

\title{Finding a clean process $B^- \to K^- D^0 K^0$ to probe absolutely exotic four-quark states}

\author{Man-Yu Duan$^{1,2}$}\email{duanmy@ppsuc.edu.cn}
\author{Guan-Ying Wang$^{3}$}\email{wangguanying@henu.edu.cn}
\author{Yun Liang$^{4}$}
\author{En Wang$^{5}$}\email{wangen@zzu.edu.cn}
\author{Xiang Liu$^{6,7,8,9}$}\email{xiangliu@lzu.edu.cn}
\author{Dian-Yong Chen$^{2,7}$\footnote{Corresponding author}}\email{chendy@seu.edu.cn}

\affiliation{$^1$School of Information Network Security, People's Public Security University of China, Beijing 100038, China\\
$^2$School of Physics, Southeast University, Nanjing 210094, China\\
$^3$Joint Research Center for Theoretical Physics, School of Physics and Electronics, Henan University, Kaifeng 475004, China\\
$^4$Department of Basics, Officers College of PAP, Chengdu 610213, China\\
$^5$School of Physics, Zhengzhou University, Zhengzhou 450001, China\\
$^6$School of Physical Science and Technology, Lanzhou University, Lanzhou 730000, China\\
$^7$Lanzhou Center for Theoretical Physics, Lanzhou University, Lanzhou 730000,China\\
$^8$MoE Frontiers Science Center for Rare Isotopes, Lanzhou University, Lanzhou, Gansu 730000, China\\
$^9$Research Center for Hadron and CSR Physics, Lanzhou University and Institute of Modern Physics of CAS, Lanzhou 730000, China
}

\begin{abstract}

Motivated by the observations of $T_{c\bar{s}0}(2900)^0$ and $T_{c\bar{s}0}(2900)^{++}$, we propose to search for $\tcsbar^0$ in the cleaner process $B^- \to K^- D^0 K^0$. In the $D^*K^*$ molecular picture, our estimates suggest that $T_{c\bar{s}0}(2900)^0$ should contribute significantly to the $D^0 K^0$ invariant mass distribution in $B^- \to K^- D^0 K^0$, as reported by the Belle II Collaboration. The corresponding fit fraction is estimated to be $(9.72\pm 3.92)\%$ or $(7.09\pm 5.88)\%$ in different fitting schemes. Further precise measurements of this process at Belle II and LHCb could be helpful for clarifying the nature of $T_{c\bar{s}0}(2900)$.

\end{abstract}

\date{\today}

\maketitle

\noindent{{\it Introduction}.---}In 2022, the LHCb Collaboration reported two new resonances, $T_{c\bar{s}0}(2900)^0$ and $T_{c\bar{s}0}(2900)^{++}$, in the $D_s^+	 \pi^-$ and $D_s^+ \pi^+$ invariant mass distributions, respectively~\cite{LHCb:2022sfr,LHCb:2022lzp}. Their measured masses and widths are~\cite{LHCb:2022sfr,LHCb:2022lzp}
\begin{eqnarray}
m_{T_{c\bar{s}0}(2900)^{0}}&=&(2892\pm14\pm15)~\mathrm{MeV}\ ,\nonumber\\
\Gamma_{T_{c\bar{s}0}(2900)^{0}}&=&(119\pm26\pm13)~\mathrm{MeV}\ ,\nonumber\\
m_{T_{c\bar{s}0}(2900)^{++}}&=&(2921\pm17\pm20)~\mathrm{MeV}\ ,\nonumber\\
\Gamma_{T_{c\bar{s}0}(2900)^{++}}&=&(137\pm32\pm17)~\mathrm{MeV}\ ,\nonumber
\end{eqnarray}
and are consistent with an isospin triplet interpretation. 
Considering isospin symmetry, the average mass and width are~\cite{LHCb:2022sfr,LHCb:2022lzp}
\begin{eqnarray}
m_{T_{c\bar{s}0}(2900)}&=&(2908\pm11\pm20)~\mathrm{MeV}\ ,\nonumber\\
\Gamma_{T_{c\bar{s}0}(2900)}&=&(136\pm23\pm13)~\mathrm{MeV}\ .\nonumber
\end{eqnarray}
Moreover, the amplitude analysis in Ref.~\cite{LHCb:2022lzp} indicate the  spin-parity quantum numbers of $T_{c\bar{s}0}(2900)$ are $J^P=0^+$. 

The quark content ($c\bar{s}q\bar{q}$) suggests $T_{c\bar{s}0}(2900)$ could be a compact tetraquark. This interpretation is supported by various theoretical approaches, including QCD sum rules~\cite{Yang:2023evp}, non-relativistic quark models~\cite{Liu:2022hbk}, the multiquark color flux tube model~\cite{Wei:2022wtr}, and flavor symmetry analyses~\cite{Dmitrasinovic:2023eei}. Production and decay properties within this picture have also been investigated~\cite{Jiang:2023rcn,Lian:2023cgs}.

Alternatively, the proximity of its mass to the $D^*K^*$ threshold makes it a candidate for a molecular state. Studies within the one-boson-exchange model~\cite{Chen:2022svh} and hidden local symmetry formalism~\cite{Duan:2023lcj} find attractive interactions capable of binding such a system, consistent with the observed state. Molecular interpretations have been used to predict its decay properties~\cite{Yue:2022mnf,Agaev:2022eyk}, although some frameworks suggest a bound state is unlikely~\cite{Ke:2022ocs}. Threshold effects have also been proposed as an origin~\cite{Ge:2022dsp}.

Given these competing interpretations, identifying $T_{c\bar{s}0}(2900)$ in cleaner production channels is crucial. Its dominant predicted decay to $DK$~\cite{Yue:2022mnf} makes $B$ decays a promising environment. While $T_{c\bar{s}0}(2900)^{++}$ was proposed to contribute to $B^+ \to D^+ D^- K^+$~\cite{Duan:2023qsg}, and some experimental evidence exists~\cite{LHCb:2020pxc}, this channel is complicated by numerous overlapping resonances (e.g., higher charmonia and the isoscalar $X_{0,1}(2900)$ states) and their interference with nonresonant backgrounds~\cite{LHCb:2020pxc,LHCb:2020bls}.

We therefore propose the decay $B^- \to K^- D^0 K^0$ as a cleaner probe for the neutral partner, $T_{c\bar{s}0}(2900)^0$. This channel offers distinct advantages: (i) the $D^0 K^0$ spectrum receives contributions only from possible isovector states like $T_{c\bar{s}0}(2900)^0$, with no background from conventional charmed-strange mesons; (ii) the isoscalar $X_{0,1}(2900)$ states do not contribute; and (iii) charmonium backgrounds are absent, as the final state contains only one charmed meson. The measured branching fraction, $(5.5 \pm 1.6) \times 10^{-4}$~\cite{ParticleDataGroup:2024cfk}, is sufficiently large for experimental study. In this work, we investigate the production of $T_{c\bar{s}0}(2900)^0$ in $B^- \to K^- D^0 K^0$ within the molecular scenario.


\begin{figure}[thb]
  \centering
\subfigure{\includegraphics[scale=0.4]{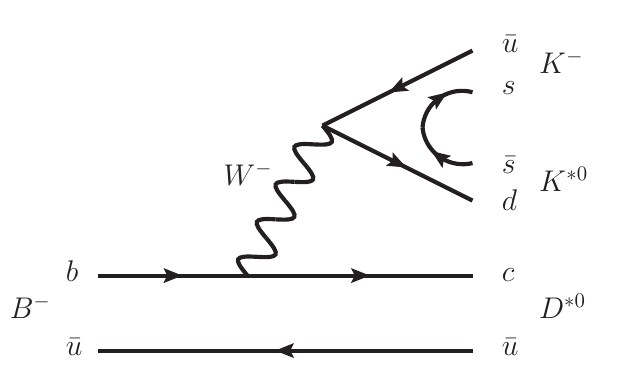}\label{KDstarKstar.pdf}\put(-70,-10){(a)} }  \subfigure{\includegraphics[scale=0.4]{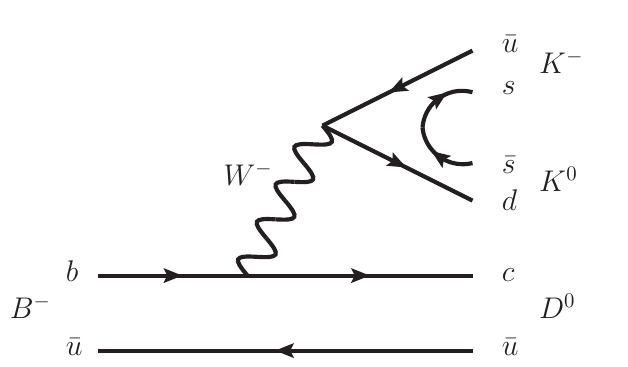}\label{KDK.pdf}\put(-70,-10){(b)} }
  \caption{Diagrammatic decay at the quark level for (a) $B^- \to K^- D^{*0} K^{*0}$ and (b) $\proc$ processes.}
  \label{Fig:quarklevel}
  \end{figure}


\noindent{{\it Contribution from $\tcsbar^0$}.---}In this work, we treat $\tcsbar^0$ as an $S$-wave molecular state of $D^{\ast 0} K^{\ast 0}$,
\begin{eqnarray}
\left|T_{c\bar{s}0}(2900)^0\right\rangle = \left|{D}^{*0} {K}^{*0}\right\rangle .\nonumber
\end{eqnarray}
which couples dominantly to its constituents. The measured branching fraction $\mathcal{B}(B^-\to K^- D^{\ast0} K^{\ast0}) = (1.5 \pm 0.4) \times 10^{-4}$~\cite{ParticleDataGroup:2024cfk} suggests that $B^-\to K^- \tcsbar^0$ can proceed via $D^{\ast0} K^{\ast0}$ rescattering.
  
\begin{figure}[t]
  \centering
  \subfigure{\includegraphics[scale=0.46]{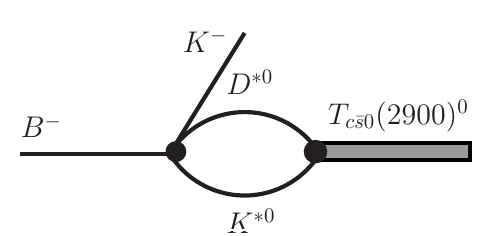}\label{res.pdf}\put(-55,-10){(a)} }
  \subfigure{\includegraphics[scale=0.46]{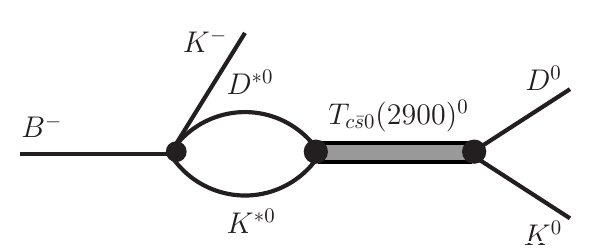}\label{resKDK.pdf}\put(-65,-10){(b)} }
  \caption{A sketch diagram of (a) the production of $\tcsbar^0$ in the $D^{*0} K^{*0}$ molecular frame and (b) further decay to $D^0 K^0$.}
  \label{Fig:Mec}
  \end{figure}

At the quark level, $B^-$ decays via external $W^-$ emission [Fig.~\ref{Fig:quarklevel}(a)]. The $b$ quark transitions into a $c$ quark and a $W^-$, which subsequently decays into $\bar{u} d$. The $c$ quark and the spectator $\bar{u}$ from the $B^-$ form a $D^{*0}$, while the $\bar{u}$, $d$ from the $W^-$ together with an $s\bar{s}$ pair from the vacuum hadronize into $K^-$ and $K^{*0}$. The $S$-wave transition amplitude, matching the angular momentum of the $B^-$, is
\begin{eqnarray}
-i t^\prime= -i C_1 {\bm \epsilon (D^{*0} ) } \cdot {\bm \epsilon (K^{*0} ) }
\label{eq:tildet},
\end{eqnarray}
where $\epsilon$ denotes the polarization vector of the vector meson, with the standard nonrelativistic normalization $\sum \epsilon_i\epsilon^*_j=\delta_{ij}$. The constant $C_1$ is a pure normalization factor encompassing the overall coupling and phase-space factors of the $B^- \to K^- D^{*0} K^{*0}$ process, and it is eventually absorbed into the fitting parameter $a_4$ in Eq.~\eqref{tcsbara4} with no independent physical implication. The $D^{0}$ and $K^{0}$ then couple to the $I(J^P)=1(0^+)$ molecular state $\tcsbar^0$ as shown in Fig.~\ref{Fig:Mec}(a).

Following Refs.~\cite{Molina:2008jw, Liang:2010ddf}, the vertex for a resonance $R_J$ with spin $J$ decaying to $D^{0} K^{0}$ can be projected as
\begin{eqnarray}
\mathcal{V}^{(0)}&=& \frac{1}{3} \epsilon_i(D^{*0}) \epsilon_j(K^{*0}) \delta_{ij} , \nonumber \\
\mathcal{V}^{(1)}&=& \frac{1}{2} \left [ \epsilon_i(D^{*0}) \epsilon_j(K^{*0}) -\epsilon_j(D^{*0}) \epsilon_i(K^{*0}) \right ] ,  \nonumber  \\
\mathcal{V}^{(2)}&=& \frac{1}{2} \left [ \epsilon_i(D^{*0}) \epsilon_j(K^{*0}) +\epsilon_j(D^{*0}) \epsilon_i(K^{*0}) \right ]   \nonumber \\
& &- \frac{1}{3} \epsilon_l(D^{*0}) \epsilon_l(K^{*0}) \delta_{ij}.\nonumber
\label{Eq:Project}
\end{eqnarray}
For the scalar ($J=0$) case, the amplitude for $B^- \to K^- \tcsbar^0$ [Fig.~\ref{Fig:Mec}(a)] becomes
\begin{eqnarray}
-i t_{2a} &=& -i  C_1 \epsilon_{\alpha} (D^{*0} )  \epsilon_{\beta} (K^{*0} ) \delta^{\alpha \beta} G\Big(M_{D^0 K^0}, m_{D^{*0}}, m_{K^{*0}}\Big)  \nonumber \\
& & \times  \frac{1}{3} \epsilon_l^* (D^{*0} )  \epsilon_l^* (K^{*0} ) \delta_{ij} g_{T^{0}_{c\bar{s}0}D^{*0} K^{*0}}  \nonumber \\
&=& -i C_1 \delta_{ij} G\Big(M_{D^0 K^0}, m_{D^{*0}}, m_{K^{*0}}\Big) g_{T^{0}_{c\bar{s}0}D^{*0} K^{*0}},\label{Eq:t2a}
\end{eqnarray}
where $G\Big(M_{D^0 K^0}, m_{D^{*0}}, m_{K^{*0}}\Big)$ is defined explicitly in the following section, we used $\sum_{\rm pol} \epsilon_i (R) \epsilon_j^* (R) = \delta_{ij}$ ($R = D^{0}, K^{0}$) and $\sum_{ij} |\delta_{ij}|^2 = 3$. The function $G(M_{D^0 K^0}, m_{D^{0}}, m_{K^{0}})$ is the two-meson loop integral for the $D^{0} K^{0}$ intermediate state.

Including the subsequent decay $\tcsbar^0 \to \dzkz$ [Fig.~\ref{Fig:Mec}(b)], the full cascade amplitude for $B^- \to K^- \tcsbar^{0} \to K^- \dzkz$ is
\begin{eqnarray}
-it_{2b} &=& -i C_1 \delta_{ij} G\Big(M_{\dzkz}, m_{D^{*0}}, m_{K^{*0}}\Big) \nonumber\\ &&\times \frac{g_{T^{0}_{c\bar{s}0}D^{*0} K^{*0}} g_{T^{0}_{c\bar{s}0}\dzkz}}{M^2_{\dzkz} -m^2_{T^{0}_{c\bar{s}0}}+i m_{T^{0}_{c\bar{s}0}}\Gamma_{T^{0}_{c\bar{s}0}}} ,\label{eq:t2b}
\end{eqnarray}
with $M^2{\dzkz}=(P_{D^0}+P_{K^0})^2$. Its modulus squared is
\begin{eqnarray}
\sum |t_{2b}|^2 &=& 3 C_1^2 \  \Big|G\Big(M_{\dzkz}, m_{D^{*0}}, m_{K^{*0}}\Big)\Big|^2\     \nonumber\\
& &\times \left|\frac{g_{T^{0}_{c\bar{s}0} D^{*0} K^{*0}} g_{T^{0}_{c\bar{s}0} \dzkz}}{M^2_{\dzkz} -m^2_{T^{0}_{c\bar{s}0}}+i m_{T^{0}_{c\bar{s}0}}\Gamma_{T^{0}_{c\bar{s}0}}}\right|^2.
\end{eqnarray}

The loop function is defined as
\begin{eqnarray}
&&G(\sqrt{s},m_1,m_2)
\nonumber\\
&& =i\int\frac{d^4q}{(2\pi)^4}\frac{1}{\Big(q^2-m_1^2+i\epsilon\Big)\Big((q-P)^2-m_2^2+i\epsilon\Big)}\ ,\nonumber
\label{eq:loopex}
\end{eqnarray}
where $m_1$ and $m_2$ are the meson masses, $q$ is the four-momentum of one meson in the centre-of-mass frame, and $P$ is the total four-momentum of the two-meson system ($s=P^2=M^2_{D^0K^0}$).

We evaluate $G$ using dimensional regularization~\cite{Oller:2000fj,Duan:2020vye}:
\begin{eqnarray}
G'(\sqrt{s},m_1,m_2)&=&\frac{1}{16\pi^2}\left[\alpha+\log\frac{m_1^2}{\mu^2}+\frac{m_2^2-m_1^2+s}{2s}\log\frac{m_2^2}{m_1^2}\right. \nonumber\\
&&+\frac{|\vec{q}\,|}{\sqrt{s}}\left(\log\frac{s-m_2^2+m_1^2+2|\vec{q}\,|\sqrt{s}}{-s+m_2^2-m_1^2+2|\vec{q}\,|\sqrt{s}}\right. \nonumber \\
&&+\left. \left. \log\frac{s+m_2^2-m_1^2+2|\vec{q}\,|\sqrt{s}}{-s-m_2^2+m_1^2+2|\vec{q}\,|\sqrt{s}}\right)\right] ,
\label{eq:loopexdm}
\end{eqnarray}
with
\begin{equation}
|\vec{q}\,|=\frac{\sqrt{\left[s-(m_1+m_2)^2\right]\left[s-(m_1-m_2)^2\right]}}{2\sqrt{s}}.\nonumber
\end{equation}
The parameters $\mu=1500$~MeV and $\alpha=-1.474$ are taken from studies of the $D^*K^*$ interaction~\cite{Lyu:2023aqn,Lyu:2023ppb,Duan:2023qsg}, consistent with those used for $D^*\bar{K}^*$ in Refs.~\cite{Dai:2022qwh, Dai:2022htx}. It should be noted that the values of these parameters only slight change the lineshape of $|G^\prime(\sqrt{s},m_1,m_2)|$, while its overall magnitude could be absorbed by the value of $C_1$.

To account for the finite width of the $K^\ast$, we convolve the loop function with the $K^\ast$ spectral function~\cite{Geng:2006yb,Wang:2019mph,Ding:2023eps,Ding:2024lqk}:
\begin{equation}
F(\sqrt{s}_{K^\ast})=-\frac{1}{\pi}\mathrm{Im}\left\{\frac{1}{s_{K^\ast}-m^2_{K^\ast}+im_{K^\ast}\Gamma_{K^\ast}}\right\},
\end{equation}
so that,
\begin{eqnarray}
&&G(\sqrt{s},m_{K^\ast},m_{D^\ast}) \nonumber\\ &&\qquad =\frac{\displaystyle\int^{(m_{K^\ast}+2\Gamma_{K^\ast})^2}_{(m_{K^\ast}-2\Gamma_{K^\ast})^2}ds_{K^\ast}G'(\sqrt{s},\sqrt{s}_{K^\ast},m_{D^\ast}) F(\sqrt{s_{K^\ast}})}{\displaystyle\int^{(m_{K^\ast}+2\Gamma_{K^\ast})^2}_{(m_{K^\ast}-2\Gamma_{K^\ast})^2}ds_{K^\ast}F(\sqrt{s_{K^\ast}})}, \qquad \ 
\label{eq:loopgamma}
\end{eqnarray}
The integration interval $(m_{K^\ast}\pm 2\Gamma_{K^\ast})^2$ adequately covers the $K^\ast$ resonance region.

The coupling $g_{T_{c\bar{s}0}^0 D^{0} K^{0}}$ is obtained from the compositeness condition~\cite{Weinberg:1965zz, Baru:2003qq,Albaladejo:2022sux, Wu:2023fyh}:
\begin{equation}
g_{T_{c\bar{s}0}^0 D^{*0} K^{*0}}^2=16 \pi (m_{D^{*0}}+m_{K^{*0}})^2 \tilde{\lambda}^2 \sqrt{\frac{2 \Delta E}{\mu}} ,
\label{eq:g1}
\end{equation}
where  the compositeness parameter $\tilde{\lambda}$ is a standard normalization factor in the hadronic molecular picture, where $\tilde{\lambda}=1$ strictly corresponds to a pure $D^{*0} K^{*0}$ molecular configuration of $T_{c\bar{s}0}(2900)^0$ (no compact tetraquark component), consistent with the core assumption of our work.  $\Delta E = m_{D^{*0}}+m_{K^{*0}} - m_{T_{c\bar{s}0}^0}$ is the binding energy, and $\mu = m_{D^{0}} m_{K^{0}} / (m_{D^{*0}}+m_{K^{*0}})$ is the reduced mass.

The coupling $g_{T_{c\bar{s}0}^0 D^{0} K^{0}}$ is estimated from the partial width $\Gamma_{T_{c\bar{s}0}^0 \to {D}^0 {K}^0}$. Using an effective Lagrangian approach,
\begin{eqnarray}
\Gamma_{T_{c\bar{s}0}^0 \to {D}^0 {K}^0}&=&\frac{1}{8\pi} \frac{1}{m^2_{T_{c\bar{s}0}^0}} g_{T_{{c}\bar{s}0}^0{D}^0 {K}^0}^2 q_{D^0} ,
\end{eqnarray}
with
\begin{eqnarray}
q_{{D}^0}&=&\frac{\lambda^{1/2}\left(m^2_{T_{{c}\bar{s}0}^0},m^2_{{D}^0},m^2_{{K}^0}\right)}{2 m_{T_{{c}\bar{s}0}^0}} ,\nonumber
\label{eq:coupling1}
\end{eqnarray}
where $\lambda(x,y,z)= x^2+y^2+z^2-2xy-2yz-2xz$ is the KÃ¤llÃ©n function. Ref.~\cite{Yue:2022mnf} estimates the branching fraction $\mathcal{B}(\tcsbar^0 \to {D}^0 {K}^0) \approx 60-70\%$; we take the central value of $65\%$ to determine $g_{T_{c\bar{s}0}^0 D^0 K^0}$.


 \noindent{{\it Contributions from $\rho(770)^-$, $\rho(1450)^-$ and $a_0(980)^-$}.---}In addition to the potential signal from $\tcsbar^0$, contributions may arise from resonances in the $K^- K^0$ system. The Belle Collaboration reported the $K^-K_S^0$ invariant mass distribution in $B^- \to K^- D^0 K_S^0$, showing clear near-threshold structures~\cite{Belle:2002gzj}. The underlying weak decay topology is similar to that in $\tau^- \to K^- K^0_s \nu_\tau$ (up to phase space). Early CLEO measurements suggested that these near-threshold enhancements could be qualitatively described by a $\rho$-like intermediate mechanism~\cite{CLEO:1996rit}. Recently, the Belle II Collaboration revisited $\proc$~\cite{Belle-II:2024xtf}. Their analysis, which included only the $\rho(1450)$ contribution, partially reproduces the peak position but does not fully describe the measured distributions; they note that an even-spin intermediate state cannot be excluded.

 \begin{figure}[htb]
 \centering
 \includegraphics[scale=0.65]{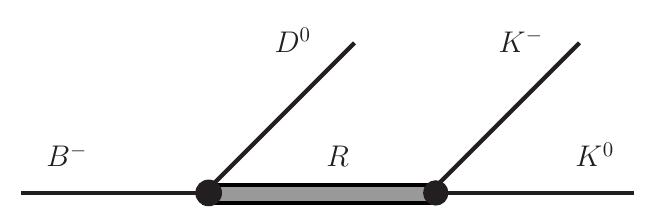}
  \caption{Feynman diagram for the contribution from $\rho(770)^-/\rho(1450)^-/a_0(980)^-$ in the $\proc$ process, the $R$ represents the resonance $\rho(770)^-/\rho(1450)^-/a_0(980)^-$.}
  \label{Fig:rho}
  \end{figure}

In this work, we consider contributions from $\rho(770)^-$, $\rho(1450)^-$, and $a_0(980)^-$. The $B^-$ decays to $D^0\rho(770)^-/\rho(1450)^-$ in a $P$-wave, with the $\rho$ states subsequently coupling to $K^- K^0$ also in $P$-wave. For the $a_0(980)^-$, both the production $B^- \to D^0 a_0(980)^-$ and the decay $a_0(980)^- \to K^- K^0$ proceed in $S$-wave. The corresponding amplitudes [Fig.~\ref{Fig:rho}] are~\cite{Wang:2020wap,Wang:2022nac}
\begin{eqnarray}
t_{\rho(770)}&=&\frac{a_1}{m_{\rho(770)}} \cdot \frac{|\vec{p}_{K^-}| \cdot |\vec{p}_{D^0}| \cdot \mathrm{cos}\,\theta } {M_{K^- K^0}-m_{\rho(770)}+i\frac{\Gamma_{\rho(770)}}{2}}  , \nonumber\\
t_{\rho(1450)}&=&\frac{a_2}{m_{\rho(1450)}} \cdot \frac{|\vec{p}_{K^-}| \cdot |\vec{p}_{D^0}| \cdot \mathrm{cos}\,\theta } {M_{K^- K^0}-m_{\rho(1450)}+i\frac{\Gamma_{\rho(1450)}}{2}}  , \nonumber\\
t_{a_0(980)}&=&a_3 \cdot \frac{m_{a_0(980)}}{M_{K^- K^0}-m_{a_0(980)}+i\frac{\Gamma_{a_0(980)}}{2}}  ,\nonumber
\end{eqnarray}
where $a_1$, $a_2$, and $a_3$ are relative strengths. Here $\vec{p}_{K^-}$ and $\vec{p}_{D^0}$ are the momenta of $K^-$ and $D^0$ in the $K^-K^0$ rest frame, and $\theta$ is the angle between them in the $K^-K^0$ frame. Explicitly~\cite{Wang:2022nac,Wang:2015pcn},
\begin{eqnarray}
|\vec{p}_{K^-}|&=&
\frac{\lambda^{1/2}\left[M^2_{K^- K^0},m^2_{K^-},m^2_{K^0}\right]}{2 M_{K^- K^0}} , \nonumber\\
|\vec{p}_{D^0}|&=&\frac{\lambda^{1/2}\left[m^2_{B^-},m^2_{D^0},M^2_{K^- K^0}\right]}{2 M_{K^- K^0}} , \\
\mathrm{cos}\,\theta&=&\frac{M^2_{D^0 K^0}-m^2_{B^-}-m^2_{K^-}+2P^0_{B^-}P^0_{K^-}}{2|\vec{p}_{K^-}||\vec{p}_{D^0}|},\nonumber
\end{eqnarray}
with $P^0_{B^-}=\sqrt{m^2_{B^-}+|\vec{p}_{D^0}|^2}$ and $P^0_{K^-}=\sqrt{m^2_{K^-}+|\vec{p}_{K^-}|^2}$ in the $K^-K^0$ rest frame.

For convenience in fitting, we recast the $\tcsbar^0$ amplitude [Eq.~\eqref{eq:t2b}] as
\begin{eqnarray}
t_{T_{c\bar{s}0}^0} =a_4 \frac{G\Big(M_{\dzkz}, m_{D^{*0}}, m_{K^{*0}}\Big) g_{T^{0}_{c\bar{s}0}D^{*0} K^{*0}} g_{T^{0}_{c\bar{s}0}\dzkz} }{ M^2_{\dzkz}-m^2_{T^{0}_{c\bar{s}0}}+im_{T^{0}_{c\bar{s}0}}\Gamma_{T^{0}_{c\bar{s}0}} } ,
\label{tcsbara4}
\end{eqnarray}
where $a_4$ absorbs the overall constant $3C_1^2$.

The total amplitude is then \cite{Lyu:2024qgc}
\begin{equation}
|t_{\mathrm{total}}|^2=|t_{T_{c\bar{s}0}^0}+t_{a_0(980)}e^{i\phi_1}+t_{\rho(770)}e^{i\phi_2}+t_{\rho(1450)}e^{i\phi_3}|^2 ,\nonumber
\end{equation}
with relative phases $\phi_{1,2,3}$. The double differential decay rates are
\begin{eqnarray}
\frac{d^2\Gamma}{dM_{\dzkz}dM_{K^- K^0}}=\frac{1}{(2\pi)^3} \frac{M_{\dzkz} M_{K^- K^0}}{8m^3_{B^-}}|t_{\mathrm{total}}|^2 , \\
\frac{d^2\Gamma}{dM_{\dzkz}dM_{D^0 K^-}}=\frac{1}{(2\pi)^3} \frac{M_{\dzkz} M_{D^0 K^-}}{8m^3_{B^-}}|t_{\mathrm{total}}|^2 ,
\label{eq:widd}
\end{eqnarray}
from which the single differential distributions $d\Gamma/dM_{\dzkz}$, $d\Gamma/dM_{K^- K^0}$, and $d\Gamma/dM_{D^0 K^-}$ are obtained by integration.


  
  \begin{figure*}[htpb]
\centering
 \includegraphics{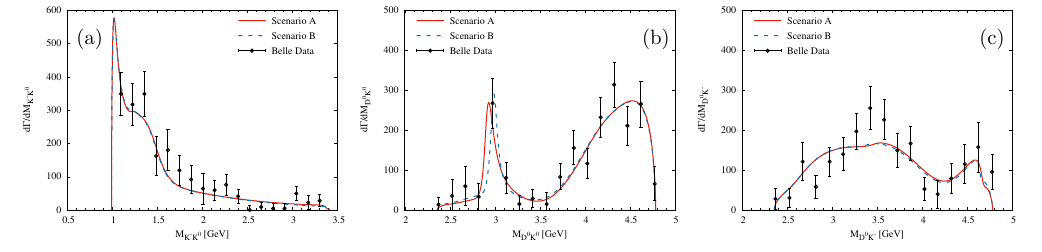}
  \caption{(a) $K^- K^0$, (b) $D^0 K^0$, (c) $D^0 K^-$ invariant mass distributions of the process $\proc$ for scenario A/B. The experimental data are taken from the first subfigure of  Figs. $9$, $13$ and $14$ of Ref.~\cite{Belle-II:2024xtf}, respectively.}
  \label{Fig:newresults}
  \end{figure*}

\renewcommand\arraystretch{1.5}
\begin{table}[t]
 \begin{center}
\caption{The values of the fitting parameters in Scenario A and B.}
 \label{table:fitvalues}
\setlength{\tabcolsep}{2 mm}{
\begin{tabular}{p{2cm}<\centering p{2.5cm}<\centering p{2.5cm}<\centering }
\toprule[0.5 pt]
\toprule[0.5 pt]
 \multirow{2}{*}{Parameters}  & \multicolumn{2}{c}{Values}\\
 \cline{2-3}
 & Scenario A & Scenario B \\
 \midrule[0.5pt]
$a_1$ & $1115.41 \pm 153.66$ &$1140.87 \pm 161.56$ \\
$a_2$ & $317.62 \pm 16.51$ & $320.22 \pm 18.58$  \\
$a_3$ & $1388.11 \pm 88.46$ & $1373.96 \pm 95.71$\\
$a_4$ & $4868.88 \pm 962.05$ & $3646.73 \pm 923.63$  \\
$\phi_1$ & $0.97 \pm 0.12$ &$0.66 \pm 0.09$\\  
$\phi_2$ & $6.25 \pm 2.60$ &$5.90 \pm 2.01$\\
$\phi_3$ & $2.95 \pm 0.43$ & $2.63 \pm 0.35$  \\
$m_{T_{c\bar{s}0}^0}$ (MeV) & ---& $2987.50 \pm 56.55$\\
$\Gamma_{T_{c\bar{s}0}^0}$ (MeV)  & ---& $94.19 \pm 66.37$\\
 \bottomrule[0.5 pt]
  \bottomrule[0.5 pt]
\end{tabular} }
  \end{center}
\end{table}

\noindent{{\it Numerical results}.---}Our analysis includes resonant contributions from $\tcsbar^0$, $\rho(770)^-$, $\rho(1450)^-$, and $a_0(980)^-$. For the masses and widths of $\rho(770)^-$, $\rho(1450)^-$, and $a_0(980)^-$ are fixed to the PDG values:  $m_{a_0(980)} = 980~\mathrm{MeV}$, $\Gamma_{a_0(980)} = 100~\mathrm{MeV}$; $m_{\rho(770)} = 775.11~\mathrm{MeV}$, $\Gamma_{\rho(770)} = 149.1~\mathrm{MeV}$; and $m_{\rho(1450)} = 1465~\mathrm{MeV}$, $\Gamma_{\rho(1450)} = 400~\mathrm{MeV}$~\cite{ParticleDataGroup:2024cfk}. The coupling strengths $a_i~(i=1,\dots,4)$ and relative phases $\phi_i~(i=1,\dots,3)$ are treated as free parameters, determined by fitting the $K^- K^0$, $D^0 K^0$, and $D^0 K^-$ invariant mass distributions from the Belle~II Collaboration~\cite{Belle-II:2024xtf}. For the mass and width of the $T_{c\bar{s}0}^0$ state, we adopt two distinct schemes: in scenario A, we fix the resonance parameters to their PDG values, namely $m_{T_{c\bar{s}0}^0} = 2892~\mathrm{MeV}$ and $\Gamma_{T_{c\bar{s}0}^0} = 119~\mathrm{MeV}$, whereas in scenario B we treat them as free parameters.

The fitted parameters for the two scenarios are summarized in Table~\ref{table:fitvalues}, yielding $\chi^2/\mathrm{d.o.f.} = 48.04/(52-7) = 1.07$ for scenario A and $\chi^2/\mathrm{d.o.f.} = 46.13/(52-9) = 1.07$ for Scenario B. This demonstrates that both scenarios describe the experimental data almost equally well. From the table, the fitted mass of the resonance is found to be slightly larger than the one of $T_{c\bar{s}0}(2900)$ reported by the LHCb Collaboration~\cite{LHCb:2022sfr,LHCb:2022lzp}, whereas the fitted width is consistent with the corresponding value for $T_{c\bar{s}0}(2900)$.  Regarding the parameters $a_i,\ (i=1-4)$ and $\phi_j,\ (j=1-3)$, their values in the two different scenarios are consistent with each other within the uncertainties. The resulting invariant mass distributions are displayed in Fig.~\ref{Fig:newresults}, along with the experimental data from Belle II Collaboration for comparison. From the figure one can find that the $K^-K^0$ and $D^0 K^-$ invariant mass distributions are nearly identical, whereas the peak structure in $D^0 K^0$ invariant mass spectrum arising from $T_{c\bar{s}0}(2900)$ exhibits slight differences. More accurate experimental data would help to provide further insight into the structure near 2.9 GeV.

\begin{figure*}[htb]
\centering
 \includegraphics{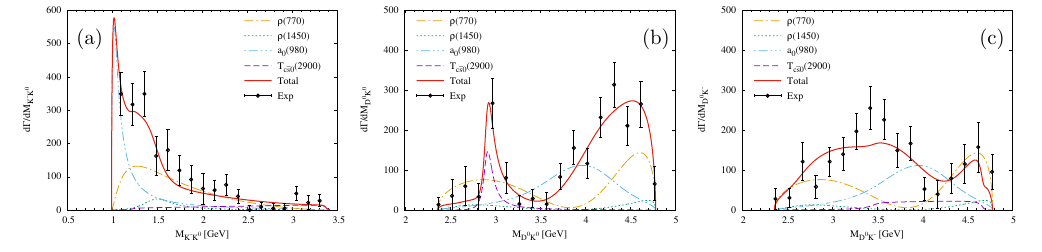}
  \caption{(a) $K^- K^0$, (b) $D^0 K^0$, (c) $D^0 K^-$ invariant mass distributions of the process $\proc$ for Scenario A. The experimental data are taken from the first subfigure of  Figs. $9$, $13$ and $14$ of Ref.~\cite{Belle-II:2024xtf}, respectively.}
  \label{Fig:app1}
  \end{figure*}

\begin{figure*}[htb]
\centering
 \includegraphics{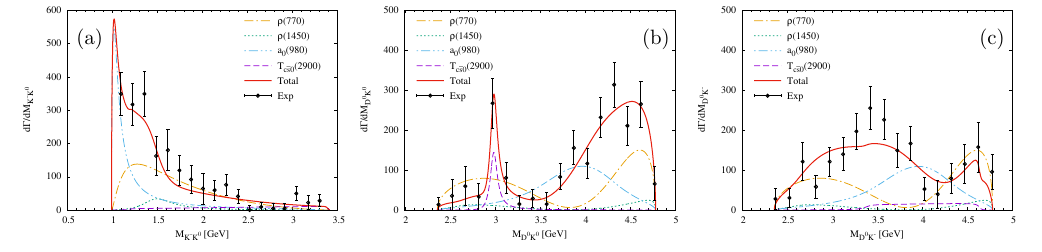}
  \caption{(a) $K^- K^0$, (b) $D^0 K^0$, (c) $D^0 K^-$ invariant mass distributions of the process $\proc$ for Scenario B. The experimental data are taken from the first subfigure of  Figs. $9$, $13$ and $14$ of Ref.~\cite{Belle-II:2024xtf}, respectively.}
  \label{Fig:app2}
  \end{figure*}

The individual contributions to the $K^- K^0$, $D^0 K^0$ and $D^0 K^-$ invariant mass distributions are displayed in Figs.~\ref{Fig:app1} and ~\ref{Fig:app2}. From the figures one can find that the near-threshold peak in $K^- K^0$ invariant mass distributions is resulted from the combined contributions of $\rho(770)^-$, $\rho(1450)^-$, and $a_0(980)^-$, while the projection from $\tcsbar^0$ is nearly flat over a wide range. In contrast, the $D^0 K^0$ distribution [Figs.~\ref{Fig:app1} (b) and ~\ref{Fig:app2} (b)] displays a clear signal for $T_{\bar{c}s0}(2900)^0$. The fit fraction for $\tcsbar^0$, defined as the ratio of the integrated intensity under its contribution (purple dashed curve) to the total (red solid curve), are $(9.72 \pm 3.92)\%$ and $(7.09 \pm 5.88)\%$ for Scenario A and B, respectively.  In the $D^0 K^-$ spectrum [Figs.~\ref{Fig:app1} (c) and ~\ref{Fig:app2} (c)], the structures arise from interference among the included resonances.


\noindent{{\it Summary}.---}In 2022, the LHCb Collaboration reported the first doubly charged tetraquark candidates, $T_{c\bar{s}0}(2900)^0$ and $T_{c\bar{s}0}(2900)^{++}$, in the $D_s^+ \pi^-$ and $D_s^+ \pi^+$ mass spectra. Their compatible resonance parameters suggest they belong to an isospin triplet. The distinctive features of these states fully open-flavor quark content and a mass near the $D^*K^*$ threshold have spurred extensive theoretical interest, with interpretations ranging from compact tetraquarks and $D^*K^*$ molecules to threshold effects.

To elucidate the nature of $T_{c\bar{s}0}(2900)$, we have proposed searching for its neutral partner in the decay $B^- \to K^- D^0 K^0$. Within the $D^*K^*$ molecular picture, our analysis shows that $\tcsbar^0$ is essential for describing the $D^0 K^0$ invariant mass distribution measured by Belle II. By including contributions from $\rho(770)^-$, $\rho(1450)^-$, $a_0(980)^-$, and $\tcsbar^0$, we achieve a consistent description of the $K^- K^0$, $D^0 K^0$, and $D^0 K^-$ spectra, with a fit fraction for $\tcsbar^0$ of $(9.72\pm 3.92)\%$ or $(7.09\pm 5.88)\%$ in different fitting schemes. Further precision measurements of this channel at Belle II and LHCb will help clarify the internal structure of the $\tcsbar$ states.

Before concluding this work, we should mention that the production process $B^- \to K^- T_{c\bar{s}0}(2900)^0$ depends on the internal structure of $T_{c\bar{s}0}(2900)^0$. The present estimation is performed entirely within the $D^\ast K^\ast$ molecular picture, and two relevant coupling constants, i.e, $g_{T_{c\bar{s}0}D^\ast K^\ast}$ and $g_{T_{c\bar{s}} DK}$, rely on this molecular assumption. The production process in the compact tetraquark scenario has not yet been investigated in the literature. Alternatively, one may distinguish between the molecular and compact tetraquark interpretations by examining the decay properties of $T_{c\bar{s}0}(2900)^0$. In the molecular frame, the estimations in Ref. \cite{Yue:2022mnf} indicate that the $T_{c\bar{s}0}(2900)$ decays dominantly in to $D K$, with a branching fraction of $60\%\sim 70\%$, which is about one order of magnitude larger than that of $D_s \pi$ channel. In the tetraquark scenario,  by contrast, the estimations in Ref.~\cite{Lian:2023cgs} suggest that the dominant decay channels of $T_{c\bar{s}0}(2900)$ are $D_s \pi$, $D K$ and $D_{s1} \pi$ with a partial widths ratio of $1:1.10:0.43$. This imply the branching fractions of $D_s \pi$ and $DK$ are of comparable magnitude. One may therefore search for $T_{c\bar{s}0}(2900)$ in the associated processes, such as, $B^- \to K^- D^0 K^0$ and $B^- \to K^- D_s^+ \pi-$, and examine the contributions of  $T_{c\bar{s}0}(2900)$ contributions in thewe channels, which would helpful to distinguish the molecular and tetraquark descriptions of this state.

\section*{Acknowledgement}
We would like to acknowledge the fruitful discussions with Xiao-Rui Lyu. This work is supported by the National Key Research and Development Program under contract No.  2024YFE0105200 and 2024YFA1610503, and the National Natural Science Foundation of China under Grant Nos. 12175037, 12475086,  12192263, and 12205075. This work is also partly supported by the Natural Science Foundation of Henan under Grand No. 232300421140, the Central Government Guidance Funds for Local Scientific and Technological Development, China (No. Guike ZY22096024), and by the Fundamental Research Funds for the Central Universities (No. 2026JKF02ZK14). X.L. is also supported by the National Natural Science Foundation of China under Grant Nos. 12335001 and 12247101, the '111 Center' under Grant No. B20063, the Natural Science Foundation of Gansu Province (No. 22JR5RA389, No. 25JRRA799), the fundamental Research Funds for the Central Universities, the project for top-notch innovative talents of Gansu province, and Lanzhou City High-Level Talent Funding.

\section*{Appendix}
In Figs.~\ref{Fig:app1} and~\ref{Fig:app2}, we display the individual contributions to the $K^-K^0$, $D^0K^0$, and $D^0 K^-$ invariant mass distributions in Scenario A and B, respectively.



\begin{thebibliography}{99}



\bibitem{LHCb:2022sfr}
R.~Aaij \textit{et al.} [LHCb],
Phys. Rev. Lett. \textbf{131}, no.4, 041902 (2023)


\bibitem{LHCb:2022lzp}
R.~Aaij \textit{et al.} [LHCb],
Phys. Rev. D \textbf{108}, no.1, 012017 (2023)





\bibitem{Yang:2023evp}
X.~S.~Yang, Q.~Xin and Z.~G.~Wang,
Int. J. Mod. Phys. A \textbf{38}, no.11, 2350056 (2023)




\bibitem{Liu:2022hbk}
F.~X.~Liu, R.~H.~Ni, X.~H.~Zhong and Q.~Zhao,
Phys. Rev. D \textbf{107}, no.9, 096020 (2023)

\bibitem{Wei:2022wtr}
J.~Wei, Y.~H.~Wang, C.~S.~An and C.~R.~Deng,
Phys. Rev. D \textbf{106}, no.9, 096023 (2022)

\bibitem{Dmitrasinovic:2023eei}
V.~Dmitra\v{s}inovi\'c,
[arXiv:2301.05471 [hep-ph]].

\bibitem{Lian:2023cgs}
D.~K.~Lian, W.~Chen, H.~X.~Chen, L.~Y.~Dai and T.~G.~Steele,
Eur. Phys. J. C \textbf{84}, no.1, 1 (2024)



\bibitem{Jiang:2023rcn}
C.~Jiang, Y.~Jin, S.~Y.~Li, Y.~R.~Liu and Z.~G.~Si,
Symmetry \textbf{15}, no.3, 695 (2023)

\bibitem{Chen:2022svh}
R.~Chen and Q.~Huang,
[arXiv:2208.10196 [hep-ph]].

\bibitem{Duan:2023lcj}
M.~Y.~Duan, M.~L.~Du, Z.~H.~Guo, E.~Wang and D.~Y.~Chen,
Phys. Rev. D \textbf{108}, no.7, 074006 (2023)

\bibitem{Yue:2022mnf}
Z.~L.~Yue, C.~J.~Xiao and D.~Y.~Chen,
Phys. Rev. D \textbf{107}, no.3, 034018 (2023)

\bibitem{Agaev:2022eyk}
S.~S.~Agaev, K.~Azizi and H.~Sundu,
Phys. Rev. D \textbf{107}, no.9, 094019 (2023)

\bibitem{Ke:2022ocs}
H.~W.~Ke, Y.~F.~Shi, X.~H.~Liu and X.~Q.~Li,
Phys. Rev. D \textbf{106}, no.11, 114032 (2022)

\bibitem{Ge:2022dsp}
Y.~H.~Ge, X.~H.~Liu and H.~W.~Ke,
Eur. Phys. J. C \textbf{82}, no.10, 955 (2022)


\bibitem{Duan:2023qsg}
M.~Y.~Duan, E.~Wang and D.~Y.~Chen,
Eur. Phys. J. C \textbf{84}, no.7, 681 (2024)

\bibitem{LHCb:2020pxc}
R.~Aaij \textit{et al.} [LHCb],
Phys. Rev. D \textbf{102}, 112003 (2020)

\bibitem{LHCb:2020bls}
R.~Aaij \textit{et al.} [LHCb],
Phys. Rev. Lett. \textbf{125}, 242001 (2020)

\bibitem{ParticleDataGroup:2024cfk}
S.~Navas \textit{et al.} [Particle Data Group],
Phys. Rev. D \textbf{110}, no.3, 030001 (2024)

\bibitem{Molina:2008jw}
R.~Molina, D.~Nicmorus and E.~Oset,
Phys. Rev. D \textbf{78}, 114018 (2008)

\bibitem{Liang:2010ddf}
W.~H.~Liang, R.~Molina and E.~Oset,
Eur. Phys. J. A \textbf{44}, 479-486 (2010)

\bibitem{Oller:2000fj}
J.~A.~Oller and U.~G.~Meissner,
Phys. Lett. B \textbf{500}, 263-272 (2001)

\bibitem{Duan:2020vye}
M.~Y.~Duan, J.~Y.~Wang, G.~Y.~Wang, E.~Wang and D.~M.~Li,
Eur. Phys. J. C \textbf{80}, no.11, 1041 (2020)

\bibitem{Lyu:2023aqn}
W.~T.~Lyu, Y.~H.~Lyu, M.~Y.~Duan, G.~Y.~Wang, D.~Y.~Chen and E.~Wang,
Eur. Phys. J. C \textbf{85} (2025) no.2, 123



\bibitem{Lyu:2023ppb}
W.~T.~Lyu, Y.~H.~Lyu, M.~Y.~Duan, D.~M.~Li, D.~Y.~Chen and E.~Wang,
Phys. Rev. D \textbf{109}, no.1, 014008 (2024)

\bibitem{Dai:2022qwh}
L.~R.~Dai, R.~Molina and E.~Oset,
Phys. Lett. B \textbf{832}, 137219 (2022)

\bibitem{Dai:2022htx}
L.~R.~Dai, R.~Molina and E.~Oset,
Phys. Rev. D \textbf{105}, no.9, 096022 (2022)

\bibitem{Geng:2006yb}
L.~S.~Geng, E.~Oset, L.~Roca and J.~A.~Oller,
Phys. Rev. D \textbf{75}, 014017 (2007)

\bibitem{Wang:2019mph}
G.~Y.~Wang, L.~Roca and E.~Oset,
Phys. Rev. D \textbf{100}, no.7, 074018 (2019)

\bibitem{Ding:2023eps}
Y.~Ding, X.~H.~Zhang, M.~Y.~Dai, E.~Wang, D.~M.~Li, L.~S.~Geng and J.~J.~Xie,
Phys. Rev. D \textbf{108}, no.11, 114004 (2023)

\bibitem{Ding:2024lqk}
Y.~Ding, E.~Wang, D.~M.~Li, L.~S.~Geng and J.~J.~Xie,
Phys. Rev. D \textbf{110}, no.1, 014032 (2024)


\bibitem{Weinberg:1965zz}
S.~Weinberg,
Phys. Rev. \textbf{137}, B672-B678 (1965)

\bibitem{Baru:2003qq}
V.~Baru, J.~Haidenbauer, C.~Hanhart, Y.~Kalashnikova and A.~E.~Kudryavtsev,
Phys. Lett. B \textbf{586}, 53-61 (2004)

\bibitem{Albaladejo:2022sux}
M.~Albaladejo and J.~Nieves,
Eur. Phys. J. C \textbf{82}, no.8, 724 (2022)


\bibitem{Wu:2023fyh}
Q.~Wu, Y.~K.~Chen, G.~Li, S.~D.~Liu and D.~Y.~Chen,
Phys. Rev. D \textbf{107}, no.5, 054044 (2023)

\bibitem{Belle:2002gzj}
A.~Drutskoy \textit{et al.} [Belle],
Phys. Lett. B \textbf{542}, 171-182 (2002)

\bibitem{CLEO:1996rit}
T.~E.~Coan \textit{et al.} [CLEO],
Phys. Rev. D \textbf{53}, 6037-6053 (1996)

\bibitem{Belle-II:2024xtf}
I.~Adachi \textit{et al.} [Belle-II],
JHEP \textbf{08}, 206 (2024)

\bibitem{Wang:2020wap}
G.~Y.~Wang, M.~Y.~Duan, E.~Wang and D.~M.~Li,
Phys. Rev. D \textbf{102}, no.3, 036003 (2020)

\bibitem{Wang:2022nac}
G.~Y.~Wang, N.~C.~Wei, H.~M.~Yang, E.~Wang, L.~S.~Geng and J.~J.~Xie,
Phys. Rev. D \textbf{106}, no.5, 056001 (2022)

\bibitem{Wang:2015pcn}
E.~Wang, H.~X.~Chen, L.~S.~Geng, D.~M.~Li and E.~Oset,
Phys. Rev. D \textbf{93}, no.9, 094001 (2016)

\bibitem{Lyu:2024qgc}
W.~T.~Lyu, S.~C.~Zhang, G.~Y.~Wang, J.~J.~Wu, E.~Wang, L.~S.~Geng and J.~J.~Xie,
Phys. Rev. D \textbf{110}, no.5, 054020 (2024)

\end{thebibliography}
\end{document}